\documentstyle[amssymb,pra,aps,manuscript]{revtex}

\input{epsf}

\begin{document}
\draft

\title{Bell Gems: the Bell basis generalized\\}

\author{ Gregg Jaeger,$^{1,2}$\\}

\bigskip

\address {$^1$College of General Studies,
Boston University}

\smallskip

\address {$^2$Quantum Imaging Laboratory,
Boston University}

\address{871 Commonwealth Ave., Boston MA 02215\\
 email: jaeger@bu.edu\\ telephone: +1 617 353-3251}

\date{\today}

\maketitle
\begin{abstract}
A class of self-similar sets of entangled quantum states is
introduced, for which a recursive definition is provided. These
sets, the ``Bell gems,'' are defined by the subsystem exchange
symmetry characteristic of the Bell states. Each Bell gem is shown
to be an orthonormal basis of maximally entangled elements.
\end{abstract}

\vfill\eject

\section{Introduction.}

The Bell basis states, the two-qubit state-vectors introduced by
Bohm [1], have proven to be the most fruitful objects of study in
the foundations of quantum mechanics and quantum information
science, due to their extreme entanglement properties [2]. The
non-factorizability that defines entanglement arises in these
states from their symmetry or antisymmetry under the binary
exchange of their qubit subsystems, each half the size of the
composite system. Here, we use these properties to define a new,
broader class of sets of quantum states, which we call ``Bell
gems,'' generalizing the Bell basis to larger qubit numbers. We
recursively define the members of this class and exhibit maps
through which they can be formally constructed. We show that the
elements of these sets form orthonormal bases for the
multiple-qubit Hilbert spaces in which they reside and possess
maximal entanglement according to the $N$-tangle measure. Finally,
we identify Bell gems and elements that have proven useful in
quantum information processing applications and provide quantum
circuits for their creation from computational basis elements.

The Bell basis for the two-qubit quantum state space ${\cal
H}_4={\cal C}^2\otimes{\cal C}^2$ is the set consisting of the
following two-qubit state vectors, the Bell states:
\begin{eqnarray}
|\Phi^\pm\rangle&=&{1\over\sqrt{2}}(|00\rangle\pm|11\rangle)\\
|\Psi^\pm\rangle&=&{1\over\sqrt{2}}(|01\rangle\pm|10\rangle)\ .
\end{eqnarray}
These states possess subsystem exchange symmetry, being of either
of the two basic forms
\begin{eqnarray}
{1\over\sqrt{2}}& &(|i\rangle|i\rangle\pm|j\rangle|j\rangle)\\
{1\over\sqrt{2}}& &(|i\rangle|j\rangle\pm|j\rangle|i\rangle)\ ,
\end{eqnarray}
where $|i\rangle$ and $|j\rangle$ are orthogonal and normalized
quantum state vectors of the same dimensionality (the two
orthogonal single-qubit states $|0\rangle$ and $|1\rangle$ of the
quantum computational basis). The forms, given in expressions (3)
and (4), of the Bell states guarantee their non-factorizability
and maximal entanglement  as measured by the concurrence or
equivalently by its square, the tangle $\tau$ [3].

Here, we construct bases generalizing the Bell basis for higher
multiparticle systems by retaining these forms as qubit number is
scaled up.

\section{Definitions and Theorems.}

The binary subsystem exchange symmetry inherent in the forms given
by expressions (3) and (4) is now incorporated in the definition
of the sets of states generalizing the Bell basis, which are
thereby retained under changes of scale. The Bell gems are
recursively defined, as follows.

\smallskip

\noindent{\bf Definition:} A {\it Bell gem}, ${\mathcal G}_d$, is
a set of mutually orthogonal normalized quantum state vectors,
lying in the $d=2^{2^n}$-dimensional Hilbert space ${\mathcal
C}^{2^{\otimes 2^n}}$, possessing the form
\begin{eqnarray}
{1\over\sqrt{2}}& &(|i\rangle|i\rangle\pm|j\rangle|j\rangle)\\
{1\over\sqrt{2}}& &(|i\rangle|j\rangle\pm|j\rangle|i\rangle)\ ,
\end{eqnarray}
where $|i\rangle\neq|j\rangle$ are elements of a Bell gem
${\mathcal G}_{d'}$ of dimensionality $d'=2^{2^{(n-1)}}$, $n\in
{\mathbb N}, n\geq 2$, the simplest Bell gem being the Bell basis,
${\mathcal G}_4$, for ${\mathcal C}^4$:
\begin{eqnarray}
|\Phi^\pm\rangle&=&{1\over\sqrt{2}}(|00\rangle\pm|11\rangle)\\
|\Psi^\pm\rangle&=&{1\over\sqrt{2}}(|01\rangle\pm|10\rangle)\ .
\end{eqnarray}

With this definition, the following theorem is seen to hold.

\noindent{\bf Theorem 1:} The Bell gem ${\mathcal G}_{2^{2^n}}$ is
an orthonormal basis for the $2^{2^n}$-dimensional Hilbert space
of quantum state vectors, ${\cal H}_{2^{2^n}}={\cal C}^{2^{\otimes
2^n}}$, that is, for the Hilbert space of $2^n$ qubits.

{\it Proof.} A Bell gem is, by the above definition, a set of
normalized linearly independent quantum state vectors of equal
qubit number. To prove this set is an orthonormal basis, it
therefore remains only show that ${\mathcal G}_{2^{2^n}}$ spans
the Hilbert space of $2^n$ qubits. That is, we now need only show
that it consists of $2^{2^n}$ elements. This can be done by
counting the number of elements of ${\mathcal G}_{2^{2^n}}$ of
each of the two available forms, given by expressions (5) and (6).
There are $2^{2^{(n-1)}}$ such linearly independent vectors of the
form (5) since each product $|k\rangle|k\rangle$ of two identical
copies of any one of the $2^{2^{(n-1)}}$ state vectors
$|k\rangle\in {\mathcal G}_{2^{2^{(n-1)}}}$ from which they are
constructed can appear only once in expression (5) (recalling
that, by definition, they are mutually orthogonal). There are,
similarly, $2^{2^{(n-1)}}(2^{2^{(n-1)}}-1)$ vectors of the form of
expression (6), since there are $2^{2^{(n-1)}}$ choices in
${\mathcal G}_{2^{2^{(n-1)}}}$ available for $|i\rangle$ and
$2^{2^{(n-1)}}-1$ choices for $|j\rangle$ remaining after the
choice of $|i\rangle$ has been made, due to the constraint that
$|j\rangle\neq |i\rangle$. Summing the total number of state
vectors of the two forms (5) and (6), the two forms themselves
being orthogonal to one another by construction, we have
$2^{2^{(n-1)}}+2^{2^{(n-1)}}(2^{2^{(n-1)}}-1)=2^{2^n}$ linearly
independent vectors. Thus, we see that the $2^n$-qubit state space
${\cal H}_{2^{2^n}}$ is spanned by the Bell gem ${\mathcal
G}_{2^{2^n}}$. The Bell gem ${\mathcal G}_{2^{2^n}}$ is,
therefore, an orthonormal basis for the space of $2^n$ qubits.
$\Box$

\bigskip

One can view the Bell gems larger than the simplest Bell gem
(${\mathcal G}_{4}$) as arising through the action of simple maps
on products of elements of the predecessor gem. Assume we are
given two subsystems, A and B, each described by elements of a
Bell gem ${\mathcal G}_{2^{2^{(n-1)}}}$. To construct elements of
a Bell gem ${\mathcal G}_{2^{2^n}}$ of the form (5) from these,
one can use the two following sorts of maps:
\begin{eqnarray}
{\mathcal P}:
|\alpha\rangle|\beta\rangle &\rightarrow& (|\alpha\rangle|\alpha\rangle+|\beta\rangle|\beta\rangle)\\
{\mathcal N}: |\alpha\rangle|\beta\rangle &\rightarrow&
(|\alpha\rangle|\alpha\rangle-|\beta\rangle|\beta\rangle)\ ,
\end{eqnarray}
where $|\alpha\rangle$ and $|\beta\rangle$ are distinct elements
of ${\mathcal G}_{2^{2^{(n-1)}}}$. To construct elements of a Bell
gem ${\mathcal G}_{2^{2^n}}$ of the form (6) from these, one can
similarly use the maps
\begin{eqnarray}
{\mathcal S}:
|\alpha\rangle|\beta\rangle &\rightarrow&
(|\alpha\rangle|\beta\rangle+|\beta\rangle|\alpha\rangle)\\
{\mathcal A}: |\alpha\rangle|\beta\rangle &\rightarrow&
(|\alpha\rangle|\beta\rangle-|\beta\rangle|\alpha\rangle),
\end{eqnarray}
 such as have been considered in
the study of quantum entanglers. The second simplest gem,
${\mathcal G}_{16}$, which unambiguously exhibits the
multiplicities described above, is shown explicitly in Section III
below.

\bigskip

Like their simplest examplar, the Bell states, the elements of all
Bell gems are maximally entangled according to the $n$-tangle
measure, $\tau_{n}\equiv |\langle \psi |\tilde{\psi}\rangle|^2,$
where $|\tilde{\psi}\rangle =\sigma_2^{\otimes n}|\psi^*\rangle$
[4,5]. This is shown in the following theorem.

\vskip 20pt

\noindent{\bf Theorem 2:} The elements of ${\mathcal G}_{2^{2^n}}$
have maximal $2^n$-tangle.

\medskip

{\it Proof}. The effect of $|k\rangle\rightarrow |{\tilde
k}\rangle$ on each element of the simplest Bell gem, ${\mathcal
G}_4$, is simply to produce a negative sign in the case it is one
of the even parity states, $|\Phi^\pm\rangle$, and to have no
effect in the case it is one of the odd parity states,
$|\Psi^\pm\rangle$. For every successor gem, in the case of
elements of the form (5), there are identical signs produced by
the transformation $|k\rangle\rightarrow |{\tilde k}\rangle$ on
the (predecessor Bell gem element) factors of each of the two
addends both of which are, therefore, unchanged. In the case of
the form (6), the sign of the addends after the transformation is
the same, leaving the element itself unchanged up to a sign. Thus
$\tau_{2^n}=|\langle k |\tilde{k}\rangle|^2=1$, {\it i.e.} all
elements of ${\mathcal G}_{2^{2^n}}$ have maximal $2^n$-tangle.
$\Box$

\bigskip

\section{Example States and Their Application.}

The simplest Bell gem is the Bell basis, the importance of which
for quantum information science ({\it e.g.} in entangled photon
quantum cryptography [6,7]) and the foundations of quantum
mechanics ({\it e.g.}\ [1]) is well known. Here, as a further
example, the next simplest Bell gem, ${\mathcal G}_{16}$ for
$2^{2^2}=16$ qubits is written explicitly and quantum circuits for
creating them are given. The significance of its elements for
quantum information processing is mentioned and a specific
application in quantum communication is pointed out.

The 16 elements $|{\mathbf e}_i\rangle$ of the $4$-qubit Bell gem
${\mathcal G}_{2^{2^2}}$ for ${\cal H}_{16}={\cal C}^{2^{\otimes
2^2}}$ can be written

\begin{eqnarray}
 {\mathcal G}_{16}=\lbrace
 &{1\over\sqrt{2}}&(|\Phi^+\rangle|\Phi^+\rangle\pm|\Phi^-\rangle|\Phi^-\rangle),\\
 &{1\over\sqrt{2}}&(|\Psi^+\rangle|\Psi^+\rangle\pm|\Psi^-\rangle|\Psi^-\rangle),\\
 &{1\over\sqrt{2}}&(|\Phi^+\rangle|\Phi^-\rangle\pm|\Phi^-\rangle|\Phi^+\rangle),\\
 &{1\over\sqrt{2}}&(|\Psi^+\rangle|\Psi^-\rangle\pm|\Psi^-\rangle|\Psi^+\rangle),\\
 &{1\over\sqrt{2}}&(|\Psi^+\rangle|\Phi^-\rangle\pm|\Phi^-\rangle|\Psi^+\rangle),\\
 &{1\over\sqrt{2}}&(|\Phi^+\rangle|\Psi^-\rangle\pm|\Psi^-\rangle|\Phi^+\rangle),\\
 &{1\over\sqrt{2}}&(|\Phi^+\rangle|\Psi^+\rangle\pm|\Psi^+\rangle|\Phi^+\rangle),\\
 &{1\over\sqrt{2}}&(|\Psi^-\rangle|\Phi^-\rangle\pm|\Phi^-\rangle|\Psi^-\rangle) \rbrace\ .
\end{eqnarray}

It is known that simple quantum logic circuits involving only two
quantum logic gates, the c-NOT and Hadamard gates, allow one to
obtain the Bell states - forming the archetypical Bell gem - from
two unentangled qubits (see, {\it e.g} [8]). Simple quantum logic
circuits will now be exhibited that similarly create the above
Bell gem elements from four unentangled qubits - that is, from
elements of the computational basis. The states ${|\mathbf
e}_1\rangle, |{\mathbf e}_2\rangle, |{\mathbf e}_3\rangle,
|{\mathbf e}_4\rangle, |{\mathbf e}_5\rangle, |{\mathbf
e}_6\rangle, |{\mathbf e}_7\rangle, $ and $|{\mathbf e}_8\rangle,$
are obtained using the circuit shown in Figure 1 with input qubits
$|x_1\rangle , |x_2\rangle , |x_3\rangle , |x_4\rangle$, where
$(x_1,x_2,x_3,x_4)$ are specifically $(0,0,0,0), (0,0,1,1),
(0,0,0,1), (0,0,1,0), (1,0,0,0), (1,0,1,1), (1,0,0,1)$ and
$(1,0,1,0)$, respectively. Similarly, the states $ {|\mathbf
e}_{13}\rangle, |{\mathbf e}_{14}\rangle, |{\mathbf
e}_{15}\rangle,$ and $|{\mathbf e}_{16}\rangle$ are obtained using
the circuit shown in Figure 2 with input qubits $|x_1\rangle ,
|x_2\rangle , |x_3\rangle , |x_4\rangle$, where
$(x_1,x_2,x_3,x_4)$ are $(0,0,0,1), (1,1,0,1), (1,0,1,1)$ and
$(0,1,1,1)$. Finally, the states ${|\mathbf e}_9\rangle, |{\mathbf
e}_{10}\rangle, |{\mathbf e}_{11}\rangle,$ and $|{\mathbf
e}_{12}\rangle$ are obtained using the circuit shown in Figure 3
with input qubits $|x_1\rangle , |x_2\rangle , |x_3\rangle ,
|x_4\rangle$, where $(x_1,x_2,x_3,x_4)$ are $(0,0,0,1),(1,1,0,1),
(1,0,1,1)$ and $(0,1,1,1)$. While the first of these circuits is
somewhat similar to that used to produce the Bell basis from the
computational basis - like the Bell circuit, it contains only one
Hadamard gate and c-NOT gates - the other two appear significantly
different from it.

The elements of the form given in expression (5), namely the first
four of the above elements, ${|\mathbf e}_1\rangle, |{\mathbf
e}_2\rangle, |{\mathbf e}_3\rangle,$ and $|{\mathbf e}_4\rangle$,
of the forms (13) and (14), are also the codes states of the
(extended) quantum erasure channel [9]. $|{\mathbf e}_2\rangle,
|{\mathbf e}_3\rangle,$ and $|{\mathbf e}_4\rangle$ are codes
states comprising a one-error correcting detected-jump quantum
code and are basis states for a decoherence-free subspace in which
universal $4$-qubit quantum computing can be carried out [10]. As
a specific application of some of these Bell gem elements, note
that the states $|{\mathbf e}_1\rangle, |{\mathbf e}_2\rangle,
|{\mathbf e}_3\rangle,$ and $|{\mathbf e}_4\rangle$ can be used
for performing error correction in the context of quantum
communication.

In particular, the above-mentioned states provide a two-qubit to
four-qubit error-correction code capable of recovery from the loss
of one photon without the loss of a qubit [11], which is highly
beneficial since photon loss is a primary factor in the
performance of simple optical quantum memories. Specifically,
there exists an error correction protocol based on these states
that provides for a more efficient quantum memory based on optical
fiber delay-line loops [12].  Successfully carrying out this error
correction process within an optical fiber loop (see [11] to view
this circuit and [12] for a description of an optical fiber loop
memory) allows quantum errors to be continually corrected on
captured flying data qubits, improving the performance of quantum
communication lines.

\section{Conclusions.}

We introduced a class of self-similar sets of quantum states,
which we call the ``Bell gems,'' and provided a recursive
definition of these sets. Every Bell gem of finite dimensionality
was shown to be an orthonormal basis for the Hilbert space in
which it naturally resides. Furthermore, the elements of all Bell
gems of $2^n$ qubits were shown to have maximal $2^n$-tangle. A
nontrivial Bell gem was exhibited and the utility of its elements
for quantum information processing was pointed out. Quantum
circuits for the creation of its elements were provided and a
specific application of some of these for quantum communication
was pointed out. Given this utility, it seems likely that the Bell
gems of yet greater complexity will be similarly useful for
quantum information applications, such as quantum coding and
quantum error correction subspaces, given the importance of
entanglement and multipartite symmetry in such applications.

\newpage

\centerline{\Large FIGURE CAPTIONS}

\bigskip

\noindent Figure 1. Quantum circuit for creating states ${|\mathbf
e}_1\rangle, |{\mathbf e}_2\rangle, |{\mathbf e}_3\rangle,
|{\mathbf e}_4\rangle,|{\mathbf e}_5\rangle, |{\mathbf
e}_6\rangle, |{\mathbf e}_7\rangle, $ and $|{\mathbf e}_8\rangle$
from $|0,0,0,0\rangle, |0,0,1,1\rangle , |0,0,0,1\rangle,
|0,0,1,0\rangle , |1,0,0,0\rangle , |1,0,1,1\rangle ,
|1,0,0,1\rangle$ and $|1,0,1,0\rangle$ utilizing Hadamard and
c-NOT gates.

\medskip

\noindent Figure 2. Quantum circuit for creating states $
{|\mathbf e}_{13}\rangle, |{\mathbf e}_{14}\rangle, |{\mathbf
e}_{15}\rangle,$ and $|{\mathbf e}_{16}\rangle$ from input states
$|0,0,0,1\rangle , |1,1,0,1\rangle , |1,0,1,1\rangle$ and
$|0,1,1,1\rangle$ utilizing Hadamard and c-NOT gates.

\medskip

\noindent Figure 3. Quantum circuit for creating states $
{|\mathbf e}_{9}\rangle, |{\mathbf e}_{10}\rangle, |{\mathbf
e}_{11}\rangle,$ and $|{\mathbf e}_{12}\rangle$ from input states
$|0,0,0,1\rangle , |1,1,0,1\rangle , |1,0,1,1\rangle$ and
$|0,1,1,1\rangle$ utilizing Hadamard, c-NOT and controlled-phase
gates introducing a controlled phase shift of $\pi/2$.

\vfill\eject

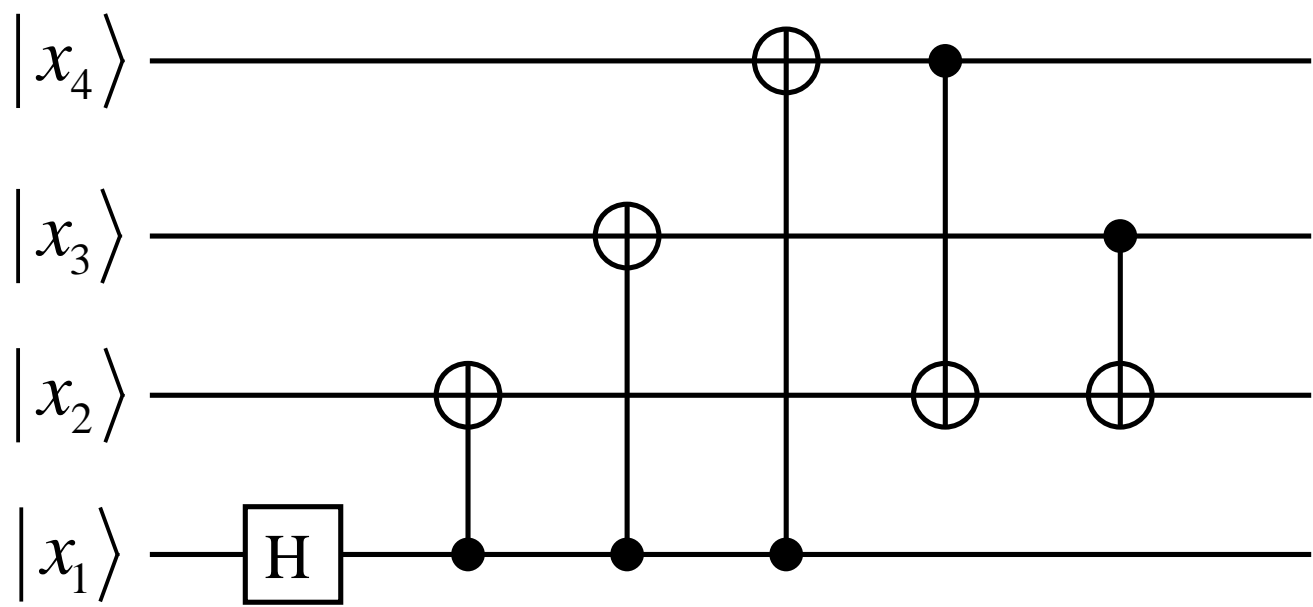
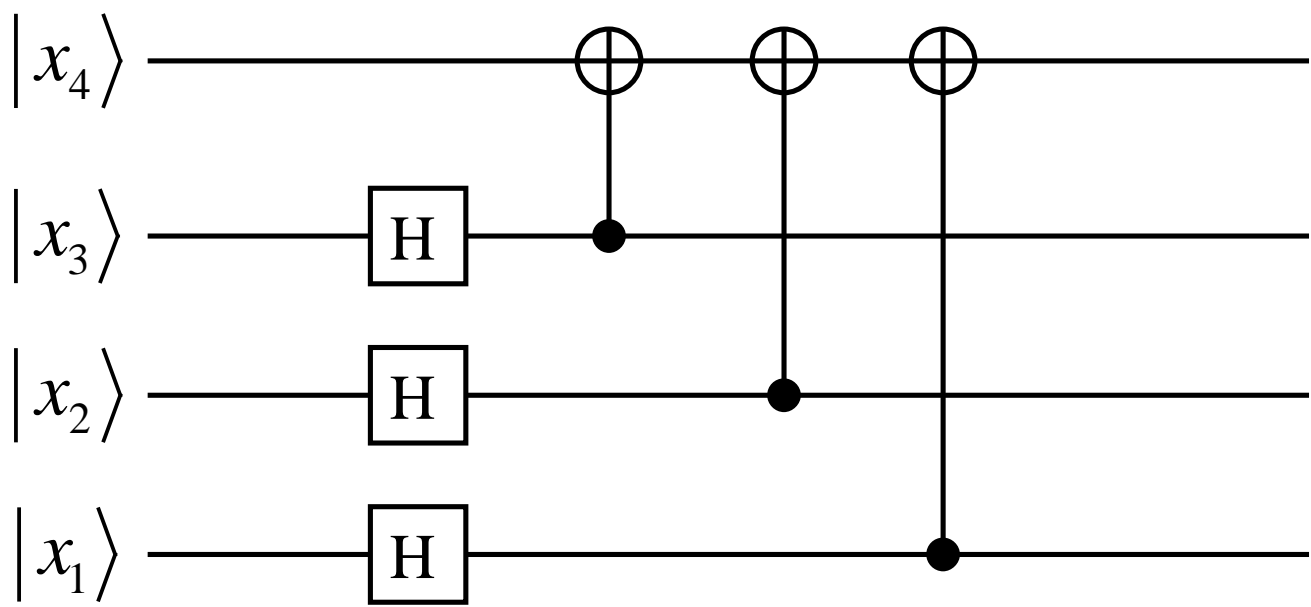
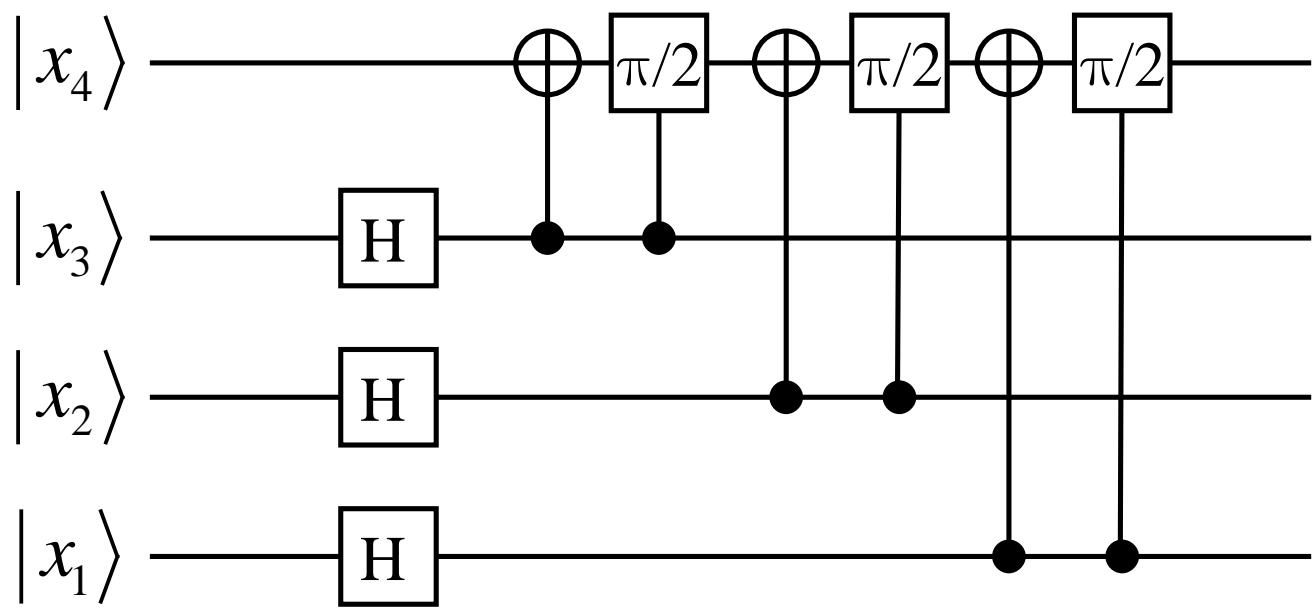
\end{document}